\documentclass[11pt]{article}
\usepackage{moriond,epsfig}

\def\be{\begin{equation}}
\def\ee{\end{equation}}
\def\bea{\begin{eqnarray}}
\def\eea{\end{eqnarray}}

\providecommand*{\ler}{\stackrel{\scriptstyle <}{\scriptstyle \sim}}

\begin{document}
\vspace*{4cm}
\title{General soft terms from Supergravity including D-terms}
\author{ Emilian Dudas$^{\star~\dagger}$ and Sudhir K Vempati$^\star$}
\address{ $^\star$Centre de Physique Theorique, \footnote{Unit{\'e} mixte
du CNRS et de l'EP, UMR 7644.}
Ecole Polytechnique-CPHT, 91128 Palaiseau Cedex, France. \\
$^\dagger$ LPT \footnote{Unit{\'e} mixte du CNRS, UMR 8627.\\
Based on talks given by S.K.V. in Moriond and Planck05.} ,
B{\^a}t. 210, Univ. de Paris-Sud, F-91405 Orsay, France.}

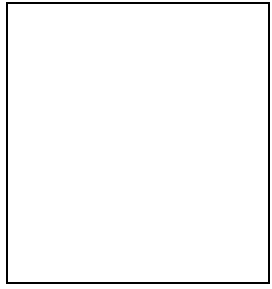
\maketitle\abstracts{
We derive general expressions for soft terms in supergravity where
D-terms contribute significantly to the supersymmetry breaking. 
Such D-terms can produce
large splitting between scalar and fermionic partners in the spectrum.
By requiring that supersymmetry breaking sets the cosmological constant to
zero, we the parameterize the soft terms when D-terms
dominate over F-terms or are comparable to them. We present
an application of our results to the split supersymmetry scenario and
briefly address the issue of moduli stabilisation. }
\section{Introduction}
%\subsection{}\label{subsec:prod}
A classical way of communicating supersymmetry breaking to the visible
sector is through gravitational interactions. In supergravity\cite{carlos},
the hidden sector scalar potential is assumed to have a minimum, preferably
generated dynamically, leading to a vacuum expectation value (vev) for
at least one of the auxiliary fields. Tree level gravitational interactions
then communicate this breaking to the visible sector generating soft terms
in the global limit. General expressions for these soft terms can then
be derived in terms of these auxiliary fields as has been pointed out
long ago by \cite{weldonsoni}. By their very nature,
such general expressions can be applied to study the soft terms in several
classes of models such as supergravity lagrangians derived from superstring
theories\cite{ibanez}.

While the existing expressions have been extremely useful, they
could be considered in a certain way as incomplete as they have
been concentrating solely on the $F$ type supersymmetry breaking
terms. It is well known that there could be $D$ type contributions
too \cite{dps}, that can arise for example in models based on anomalous $U(1)$
symmetries \cite{bd}. Furthermore, in effective lagrangians from the
Type II orientifolds with intersecting D-branes, one can expect such
D-term contributions to be naturally present. There is also a second
motivation. Recently, influenced
by the multivacuum structure of string theory\cite{bp} as possible
new view of cosmological constant problem, a new proposal has been
put forward by the authors of Ref.\cite{split}. Here,
it is proposed that the fermionic superpartners stay close
to the weak scale, whereas the scalar superpartners can be present
at scales as high as $10^9$ GeV. It would be very difficult to achieve
this kind of splitting between superpartners in supergravity models with
only $F$ type supersymmetry breaking.
Given these motivations, we present here the results
obtained in \cite{edskv} for the general expressions
for soft terms in presence of non-zero D-term contributions and study a few
applications for them. Particularly,
we sketch a model where such large D-terms can be utilised in realising 
split supersymmetry and address the issue of moduli stabilisation, of 
particular relevance for any scenario of supersymmetry breaking.
\section{General Expressions Including D-breaking}
Let us now proceed to generalise the analysis in the literature
by including abelian gauge groups $\prod_A U(1)_A$ and the corresponding
D-type contributions to the SUSY breaking. The scalar potential
now takes the form: \be \label{potential} V = e^G (G^M ~G_M
- 3) + {1 \over 2}  \sum_A g_A^2 D_A^2 , \ee
where the auxiliary $F$ fields are given by
$ G_M = {\partial G \over \partial z^M}$ and  $z$ represents the
scalar part of a chiral superfield. The index $M$ runs over all the
chiral superfields present,  matter as well as hidden sector and/or
moduli fields. At the minimum, the hidden sector auxiliary fields attain
a vev breaking supersymmetry spontaneously.
The D-terms are given by
\be \label{Dtermdef}
D_A = z^I X_{I}^A {\partial K \over \partial z^I} ~+~ \xi_A =
 {\bar z}^{\bar I} X_{I}^A {\partial K \over \partial {\bar z}^{\bar
I}}   ~+~ \xi_A \ \;;\;\; \xi_A \equiv \eta^\alpha_A \ \partial_\alpha K ,
 \label{s1} \ee
where $X_I^A$ represents the $U(1)_A$ charges of the field $\phi^I$
and $\xi_A$ denotes the Fayet-Iliopoulos terms for the abelian $U(1)$ factors.
Note that the equality between the first two  terms is a straightforward
consequence of the gauge invariance of the K\"ahler potential. We consider
the Fayet-Iliopoulos terms to be moduli dependent and we will not
explicitly discuss here the various possible mechanisms of moduli
stabilisation.
The conditions of the cancellation of the cosmological constant and the requirement of existence
of a minimum gives
\bea \label{vaccond1}
&& < e^G (G^M ~G_M - 3) + {1 \over 2}  \sum_A g_A^2 D_A^2 > \ =  0 \ , \\
\label{vaccond2a}
&& <e^G ( G^M~\nabla_K G_M  + G_K ) +  \sum_A g_A^2 D_A
(\partial_K D_A - {1 \over 2} G_K D_A ) > = 0 \ , \eea
where $\nabla$ denotes the covariant derivative on the K\"ahler manifold.
The scalar soft spectrum is defined as :
\bea
m^2_{I \bar{J}} =~ <\partial_I \partial_{\bar{J}} V>~ =
~< \nabla_{I} \nabla_{\bar{J}} V > \ , \\
m^2_{I J} =~ <\partial_I \partial_J V>~ =~ < \nabla_{I} \nabla_{J} V > \ , \\
A_{IJK} ~=~ < \nabla_{I} \nabla_{J} \nabla_{K} V > \ .
\eea We will further distinguish the visible sector (matter) fields
from those of hidden sector fields $T^{\alpha}$ (and later on flavon fields),
by requiring
$\langle G^i \rangle = 0 \quad , \quad \langle \Phi^i \rangle = 0 $, 
with $\Phi$ representing the scalar part of a matter field. Using
these we recover in the absence of D-term contributions the standard form\cite{wessbagger,ibanez} for
the soft scalar masses.
Given the form of D-terms above (\ref{s1}), we have in the vacuum, after
setting the matter fields \textit{vevs} to zero
\bea
&& \langle \partial_j D_A \rangle =
\langle \bar{v}_{\bar \beta} X^A_{\bar \beta} K_{\bar{j} \beta} +
  \eta_A^{\bar{\alpha}} K_{j \bar{\alpha}} \rangle = 0\quad , \quad  
\langle \nabla_i \nabla_j D_A \rangle = 0 \ , \nonumber \\
 \langle \partial_i \partial_{\bar j} D_A \rangle &=& K_{i {\bar j}} X_{i}^A
+ (\bar{v}^{\bar{l}} X^{\bar{l}}_A \partial_{\bar{l}} + \eta^{\bar{\alpha}}_A
\partial_{\bar{\alpha}} )~K_{i \bar{j}} \;\; , \;
\langle \nabla_i \nabla_j  \nabla_l D_A \rangle = 0
 \label{s2} \eea
The equations for the soft terms are now given by \cite{edskv}:
\bea
\label{mattscalarD} m^2_{i \bar{j}} &=& m_{3/2}^2 \left( G_{i \bar{j}}-
R_{i\bar{j}\alpha\bar{\beta}} G^\alpha G^{\bar{\beta}} \right)
+ \sum_A  g_A^2 D_A \left( X^A_i 
+  \bar{v}_{\bar{l}} X_{\bar{l}}^A \partial_{\bar{l}}
+  \eta^{\bar{\alpha}}_A \partial_{\bar{\alpha}} -
{1 \over 2}  D_A \right) G_{i \bar{j}} \ , \\
\label{mattbilinearD}
m^2_{ij} &=&  m_{3/2}^2~ \left( 2 \nabla_i G_j +~G^\alpha \nabla_i \nabla_j
G_\alpha \right) - {1 \over 2} \sum_A g_A^2 D_A^2 ( \nabla_i G_j +
{1 \over 2} g_A^2 \partial_i \partial_j f_A) 
 \ , \\
\label{mattaparamD}
 A_{ijk}&=& m_{3/2}^2 \left( 3 \nabla_i \nabla_j G_k + G^\alpha \nabla_i
   \nabla_j \nabla_k G_\alpha \right) - {1 \over 2} \nabla_i \nabla_j G_k
\sum_A g_A^2 D_A^2 \ ,
\eea
 where we have identified the gravitino mass $<e^G> = m_{3/2}^2$ and
$f_A$ is the gauge kinetic function. 
While these expressions are given for the tree
level potential, higher order corrections can play a significant role,
depending on the specifics of the model of supersymmetry breaking. In models
with small tree-level contributions, the dominant set of
corrections are of anomaly mediated type\cite{amsb1} which are
proportional to the gravitino mass $m_{3/2}$. 
A detailed analysis including various parameterisations
will be presented in \cite{edskv}.
The $\mu$ term
and the tree level gaugino mass terms are then given by \be
\label{agauginof} \mu_{ij} = m_{3/2} \ \nabla_i G_j  \quad ,
\quad M^A_{1/2} = { 1 \over 2} ( Re f_A)^{-1} m_{3/2} f_{A \alpha}
G^\alpha \ . \ee It is clear from the above analysis that in the
F-limit where D term contributions are absent,
the soft terms all typically of the same order
of magnitude without large hierarchies within themselves. These
expressions have been used to parameterise soft terms from superstring
theories as well as supergravity\cite{ibanez}.  Of course, if the
gravitino mass  is very larg $m_{3/2} >> TeV$, possible higher
derivative operators can change the pattern displayed above
completely. A consistent supergravity analysis in such a case, however,
becomes considerably more involved.

\subsection{Implications of large D-terms on the soft parameters}
Let us now systematically see the impact of the D-terms on each
of the soft parameters. As has been noted earlier, as long as they
are of the $\mathcal{O}(m_{3/2}^2)$, they do not have strong impact.
Let us now consider the limit $ m_{3/2}^2 \ler D_A \ler  m_{3/2} M_P$.\\
\noindent (i). \textit{Sfermion Mass Terms}: The most dominant contribution
to the sfermion masses from the $D$-terms are the ones which are linear
in $D$  which for $m_{3/2} \sim$ TeV
push the scalar masses to intermediate energy scale. Note that
these terms depend on the charges
of the fields under the additional $U(1)$ gauge group, thus
putting a constraint that these charges to be of definite sign. If all the
three generations of the sfermions have the same charges under the
$U(1)$ groups, this term would also be universal. \\
\noindent (ii). \textit{Higgs mass terms and the $B \mu$}: The Higgs masses
follow almost the same requirements as the soft masses. Usually, 
their charges are linked with the Giudice-Masiero mechanism\cite{gm}.
The $B \mu$ term is however special. Unlike the Higgs mass terms, 
it does not receive large contributions from D-terms, whose contributions 
can be utmost of $\mathcal{O}(m_{3/2}^2)$. If the splitting between 
the Higgs masses and the $B \mu$ is too large, it could lead to unphysical
regions in $\tan \beta$. To remedy this, alternative schemes have to be
devised, an explicit example being discussed in the next subsection.\\
\noindent (iii). \textit{A-terms}: Even if the D-terms are large, the
A-terms are typically proportional to $\mathcal{O}(m_{3/2})$. No
large enhancement is present. This is expected as A-terms break
R-symmetries. 
\subsection{A model for Split supersymmetry}
The requirement of Split supersymmetry type soft spectra are as
follows : \\
\noindent (i) Scalar soft terms :
$m_{\tilde{f}}^2~ \sim ~ \mathcal{O}(10^9 - 10^{15})$ GeV,
$(\tilde{f} ~= ~Q, ~u^c,~d^c, ~L,~e^c)$ \\
\noindent (ii). Higgs mass parameters
$m_{H_1}^2~ \sim ~m_{H_2}^2 \sim~ B_\mu ~\sim  ~
\mathcal{O}(10^9 - 10^{15})$ GeV, with one of them fine tuned to be
around the electroweak scale.\\
\noindent
(iii). The gaugino masses and the $\mu$ term are around the weak scale.\\
\noindent
From the discussion in the previous section, it was obvious that it is
just not sufficient to choose the $U(1)$ charges of the sfermions to be
positive to realise the split spectrum\footnote{see also Ref.\cite{nath}}. 
Since $B_\mu$ term does not
have large D-term contributions, we need to disentangle the $\mu$ and the
$B_{\mu}$ term by introducing (at least ) one new field $X$ and allowing a term of the
type $X H_1 H_2$ in the superpotential. If the auxiliary field $\langle F_X
\rangle \sim \langle D \rangle $, whereas $\langle X \rangle << m_{3/2}$, 
then the tuning of Higgs parameters is technically possible.
The minimal field content realising this is as
follows. The model contains an additional $U(1)$ group, with two additional
fields $X$ and $\phi$ with charges $+2$ and $-1$. The $\phi$ field
can act as a flavon field attaining a large vev close to the fundamental
scale. The superpotential and the relevant terms in the Kahler
potential, obtained by expanding in powers of the matter fields the
full supergravity , are
\bea
&& W = W_0 + W_{SSM} + \lambda_1 X H_1 H_2 + \lambda_2 X \phi^2 + \cdots \ ,
\nonumber \\
%&& K = - 3 \ ln (T + T^{\dagger}- \sum_i |\phi_i|^2) +
%{Z \over (T +T^{\dagger})^2} (\phi^{\dagger})^2 H_1 H_2 + \cdots \ .
&& K = K_0 +  \sum_{i {\bar j}} Z_{i {\bar j}} \phi^i {\bar \phi}^{\bar j}  
+ Z' (\phi^{\dagger})^2 H_1 H_2 + \cdots \ . \label{s01}
\eea
In (\ref{s01}), $W_0$ is a holomorphic function of moduli fields ,
$K_0$ is the Kahler potential for moduli, $Z_{i \bar j}$ and $Z'$ are
generically also moduli dependent and the dots denote higher order
terms in an expansion in matter fields. The main features of the model
are already captured by performing a vacuum analysis at the global
supersymmetry level. In this case, the scalar potential is given by
\be
\label{toymodel}
V = \lambda_2^2 (|\phi|^4 + 4 | X|^2 | \phi|^2) +
{ 1 \over 2} g^2 \ (2 | X|^2 - | \phi|^2 + \xi)^2 + \ldots \ ,
\ee
For $\xi >0$, the stable extremum of the above and the auxiliary fields
are given by:
\be
\langle \phi \rangle = {g^2 \over 2 \lambda_2^2 + g^2 } \xi \ , \
\langle X \rangle = 0 \ , \
\langle F_{\phi} \rangle =0 \ , \ \langle F_X \rangle =
{\lambda_2 g^2 \over 2 \lambda_2^2 + g^2 } \xi \ , \
\langle D \rangle = {2 \lambda_2^2 \over 2 \lambda_2^2 + g^2 } \xi \ .
\ee
From the above it is clear that $F_X \sim g^2 D$ and moreover are of order
of the FI term $\xi$. This is sufficient to enable the $B$ term to receive
large contributions through the term $G^X \nabla_{H_1} \nabla_{H_2} G_X$
in the eq.(\ref{mattbilinearD}). As long as $\xi$ is close to an intermediate
scale $m_{3/2}^2 << \xi \le m_{3/2} M_P$, this model seems to replicate
the hierarchical split spectrum, if one fixes the
gravitino mass at 1 TeV. However, in typical string models, the
FI term is of the $\mathcal{O}(M_{Pl}^2/16 \pi^{2})$, which would give a too
large contribution to the vacuum energy.
The correct order of magnitude could be achieved
by incorporating the above model into a higher dimensional theory. For
illustration lets us consider a 5D theory compactified over $S^1/Z_2$. The
Standard Model and the $X,~\phi$ fields live on a 3D brane, whereas the
gauge fields of the $U(1)$ are allowed to propagate in the bulk. We will
use Scherk-Schwarz mechanism to break supersymmetry.

%The modulus dependence of the Giudice-Masiero term in the Kahler
%potential is the natural one in a theory with $SL(2,Z)$ T-modular invariance.
%In the following we define $t \equiv Re \ T$.
%
%In order to suppress D-term contributions to the vacuum energy to be of the
%order $m_{3/2}^2 M_P^2$, in the following we propose a higher-dimensional
%realisation based on the Scherk-Schwarz mechanism.

The various scales in the problem are
$R = t M_5^{-1} \ , \ R M_5^3 = M_P^2 $, where
$t \equiv Re \ T$, the modulus field.
After canonically normalizing the various fields by
${\hat \phi}_i = \sqrt{t/3} \ \phi_i$ and at the global supersymmetry level,
the potential retains the form (\ref{toymodel}) with
 $\xi \sim M_5^2 = M_P^2/t $.
The four dimensional gauge coupling is given by
$g^2 = 1/t = 1/(R M_5)$, whereas the gravitino mass is given by
$m_{3/2} = { \omega / R}$, where $\omega$ is a number of order
one. The D-term contribution to the vacuum
energy is then of the form
$\langle V_D \rangle \sim g^2 M_5^4 \sim m_{3/2}^2 M_{P}^2 $,
in the right order as required by the cancellation of the vacuum
energy and realisation of the split spectrum. If the no-scale
structure is broken by the dynamics, the gauginos attain
their masses through anomaly mediation and thus we choose the gravitino
mass to be of the order of 100 TeV. In the opposite case, new sources
of gaugino masses have to be invoked, like Dirac-type masses or higher
dimensional operators if the gravitino is much heavier.
The $\mu$ term can be  generated by Giudice-Masiero mechanism and is 
$\mu \sim (v/M_5)^2 m_{3/2}$.
So, this model replicates the spectrum of the split
supersymmetry at the weak scale using large D-terms of the intermediate
scale and a 100 TeV massive gravitino.

\subsection{Moduli stabilisation problem}
As transparent in (\ref{s1}), the FI terms are field (moduli)
dependent. If no additional dynamics is present, the moduli fields
will always exbihit a runaway behaviour and the FI terms
disappear. We resume here the issue of moduli stabilisation with
realisation of large D-term contributions to soft terms discussed
in \cite{edskv}in a context similar to, but having some new
features compared the one discussed some time ago in \cite{bd}. We
would like to stress that the analysis performed in \cite{edskv}
and summarized here is also relevant for the issue of the uplift
of the energy density in the context of KKLT type moduli
stabilisation \cite{kklt}. The gauge group consists of the Standard Model
supplemented by a confining hidden sector group and an anomalous
$U(1)_X$. For simplicity we discuss the case of an supersymetric
SU(2) gauge group with one quark flavor $Q^a$ and anti-quark
${\tilde Q}^a$ where $a=1,2$ is an index in the fundamental
representation of the $SU(2)$ gauge group. The hidden sector
consists of a stack of two magnetised D9 branes in the type I
string with kinetic function $ f = S + k T$, where S is the
dilaton (super)field, T a volume (Kahler) modulus and $k$ is an
positive or negative integer determined by the magnetic fluxes in
two compact torii. The dynamical scale of the hidden sector gauge
group and the effective superpotential are 
\be \Lambda = M_P e^{- 8 \pi^2 (S+k T)/5} \ , \ 
W = W_0 + {\Lambda^5 \over M} + \lambda \varphi M \
. \label{m1} \ee In order to stabilise the modulus S we invoke the
three-form NS-NS and RR fluxes. The low energy dynamics
is described by $M =
Q^a {\tilde Q}^a$, the composite "meson" field. $W_0$ depends on
the modulus $S$ and eventually other (complex structure) moduli of
the theory and stabilise them $S=S_0$ by giving them a very large
mass. If the other relevant mass scales, the FI term and the
dynamical scale $\Lambda$ have much lower values, we can safely
integrate out these fields, by keeping the T modulus in the low
energy dynamics. Minimisation with respect to $T$ stabilises also
the Kahler modulus. Due to the anomalous nature of the $U(1)_X$,
there are mixed anomalies with the hidden sector gauge group which
translate into a chiral nature of the quark and anti-quark field,
such that the sum of their charges, equal to the $M$ meson charge,
is different from zero and, in our example, equal to $+1$.
$\varphi$ is  a field of charge $-1$ which originally participate
in the Yukawa coupling $\lambda \varphi Q^a {\tilde Q}^a$, which
plays the role of meson mass after the spontaneous symmetry
breaking of the $U(1)_X$. Along the $SU(2)$ flat directions, the
D-term scalar potential is \be V_D = {g_X^2 \over 2} \left[
(M^{\dagger} M)^{1/2} - |\varphi|^2 + k \mu^2 \right]^2 \ ,
\label{m2}\ee where $\mu$ is a mass scale determined by the T-modulus
vev. The new feature of (\ref{m2}) is that $k$ and consequently
the FI term can have both signs, whereas in the effective
heterotic string framework worked out in \cite{bd}, the FI term
had only one possible sign. In the
limit $\Lambda << \mu$, the vacuum structure and the pattern of
supersymmetry breaking in the two cases of $k$ positive and
negative are vastly different :

i) $k > 0$. In this case the vacuum can be determined as in
\cite{bd}. We find, to the lowest order in the parameter $\epsilon
\simeq (\Lambda / k^{1/2} \mu)^{5/2}$, a hierarchically small
scale of supersymmetry breaking 
\bea 
&& \langle |\varphi|^2 \rangle
= k \mu^2 \ , \ \langle M \rangle = \lambda^{-1/2} \Lambda^2 (\Lambda /
k^{1/2} \mu)^{1/2} \ , \nonumber \\
&& \langle D_X \rangle = - {\lambda \Lambda^5 \over (k^{1/2} \mu)^3} \ , \
\langle F_{\varphi} \rangle = \Lambda^2 ({ \lambda \Lambda \over k^{1/2}
\mu})^{1/2} \ , \  \langle F^{\bar M} \rangle = K^{M {\bar M}}  \partial_M W
 = - {\Lambda^5 \over k \mu^2 M_P^2} \ . 
\eea
ii) $k < 0$. In this case we find, to the lowest order in the
parameter $ \epsilon' \sim (\Lambda^2 / |k| \mu^2)^{5}$ , a large
scale of supersymmetry breaking  (for the complete expressions, see
\cite{edskv}) \be \langle |\varphi|
\rangle \sim {\Lambda^5 \over k^2 \mu^4 } \ , \ 
\langle M \rangle \sim |k| \mu^2 \ , \
\langle D_X \rangle \sim k \mu^2 \
, \ \langle F_{\varphi} \rangle \sim k \mu^2 \ , \  
\langle F^{\bar M} \rangle \sim - {\Lambda^5 \over k \mu^2 M_P^2} \ . \ee
Interestingly enough, this second case generate a large scale for 
supersymmetry breaking with large D-term contributions.
%%%%%%%%%%%%%%%%%%%%%%%%%%%%%%%%%%%%%%%%%%%%%%%%%%%%%%%%%%%%%%%%%%%%%%%%%%%%%%
\subsection{Acknowledgments}
\noindent
We would like to thank the Moriond organizers for providing a
stimulating atmosphere.
We wish to thank  S. Lavignac and Carlos Savoy for useful discussions.
\noindent
This work is supported in part by the CNRS PICS no. 2530 and 3059, INTAS grant
03-51-6346, the RTN grants MRTN-CT-2004-503369, MRTN-CT-2004-005104 and
by a European Union Excellence Grant, MEXT-CT-2003-509661.
SKV is also supported by Indo-French Centre for Promotion of
Advanced Research (CEFIPRA) project No:  2904-2 `Brane World´ Phenomenology'.

\end{document}